# Evaluating the Relationship of EV Charging Station on the Uptake of Electric Vehicles — Implication of the NEVI Formula Program

Putra Farrel Azhar[1]

**Abstract**

To achieve the federal goal to make half of all new vehicles sold in the U.S. in 2030 zero-emissions vehicles, the U.S. Department of Transportation's (DOT) Federal Highway Administration (FHWA) has employed the National Electric Vehicle Infrastructure (NEVI) Formula Program, which aims to promote the interconnected network of publicly accessible electric vehicle (EV) charging stations. By analyzing panel data on a subset of U.S. counties, this paper examines the relationship between different classes of EV charging stations towards plug-in hybrid electric vehicles (PHEVs) and battery electric vehicles (BEVs). Though an instrumental approach in identifying potential endogeneity, this study suggests that public charging stations and the implication of the NEVI formula program might not pose the optimistic benefit in achieving the federal goal. This finding indicates the need to reevaluate current strategies and explore additional incentives to enhance the adoption of EVs and the effectiveness of public charging infrastructure.

## 1   Introduction

In rural regions of the US, where 20 percent of the population and approximately 70 percent of the country's road networks are established—alternative fuel vehicles (AFVs), most notably electric vehicles (EVs), are significant in minimizing the environmental impacts of transportations compared to traditional combustion powered vehicles (US DOT, n.d). Although empirical research has highlighted the effects of federal and state incentives such as tax credits, clean vehicle rebates, and high-occupancy vehicle (HOV) lane stickers in stimulating demands for EVs, the rate for EV adoption remains 40 percent lower in rural areas (US DOT, n.d). A survey study that was conducted by McKinsey & Company in 2022 shows how 42 percent of skeptical EV buyers state that public access to charging stations is a dominant concern that impedes their decision in adoption—saying they would only consider the purchase of EVs when the availability for public charging becomes as equivalent to traditional gas

---

[1] University of California, San Diego, School of Global Policy and Strategy; Email: pazhar@ucsd.edu

stations (Fischer et al., 2024).

In light of these concerns, on November 15th, 2021, President Biden signed The Bipartisan Infrastructure Law (BIL), which allocates $1.2 trillion towards infrastructure spending and new program investments (US DOT PHMSA, 2023). Approximately $5 billion was given towards the National Electric Vehicle Infrastructure (NEVI) Formula Program, which distributes state-level funding across 2022 – 2026 to strategically construct EV charging stations and establish an interconnected network of EV charging infrastructure (California Energy Commission, n.d). Although multiple purposes are associated with the program, providing better convenience for EV adopters through a higher number of public charging stations remains the primary goal of the NEVI funding. Accordingly, this paper examines the validity of the presumed relationship that a higher number of public charging stations would provide higher EV adoptions—providing insights if an increased number of NEVI fundings and, therefore, public charging stations would help achieve the federal goal of making half of all new vehicles sold in the US in 2030 zero-emissions vehicles. The condensed policy causal result chain goes as follows:

| Inputs → | Activities → | Outputs → | Outcomes → | Final Outcomes |
|---|---|---|---|---|
| NEVI Formula Program distributes fundings to states. | States reallocates fundings toward EV charging infrastructure investments. | Increased publicly accessible EV charging stations. | Higher rate of EV charging facilities to traditional gas stations. | Increased number of EV adoptions. |

The NEVI Formula Program allocates the distribution of states' funding based on the agreed-upon EV Infrastructure Deployment Plans each state provides at the end of every fiscal year of the program's deployment (US DOT FHWA, 2024). The reallocation of funding from states to private and public entities to invest in public charging stations depends on a case-by-case scenario—adhering to the strict minimum requirements that each public charging station that received the funding be located within one mile of an established alternative fuel corridor (AFC), accommodate the DC solutions power requirement, be non-proprietary, have a minimum station power of 600 kW, to name a few. As public EV charging stations become more readily available, the presence of EV charging facilities would equate to the number of traditional gas stations and, therefore, address EV skeptics for vehicle ranges. Consequently, there will be higher rates for EV adoptions thanks to the increased proximities of public charging infrastructures. Figure 1. highlights the exponential uptake of EVs at the time of the first two rounds of NEVI funding. Although it would be irrational to conclude that NEVI had any effect on EV adoption simply from this illustration—the timing of the NEVI fundings compounded with the magnitude of growth on EV adoption makes this an intriguing case to infer if NEVI had any influence. Nevertheless, there are notable limitations that this paper remedies when directly analyzing the impact of the NEVI Formula Program on the uptake of EV adoptions. Collecting data on the financial allocation of state funding to the construction of each specific charging station proves to be a massive undertaking. There has yet to be a comprehensive public report, given that the program is still being implemented and has not been fully realized until the end of FY2026. Accordingly, EV charging stations that conform to the NEVI funding requirements will be used as the primary treatment variable for this paper's point estimate as a substitute for the impact of the county-level NEVI funding distribution. Although this approach may not allow us to directly identify the effects of public charging stations that were predominantly funded by the program due to the construction of the estimating formula, understanding the general effect of public charging stations would be the scope of this paper to infer whether funding should be expanded accordingly.

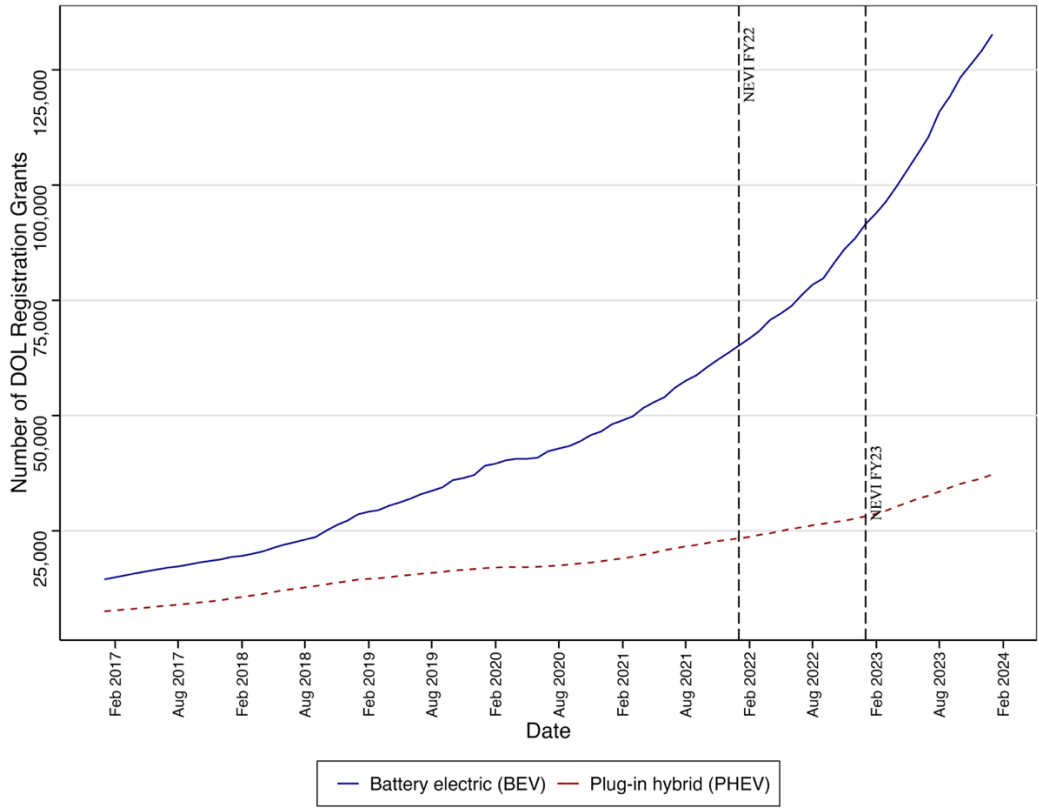

Figure 1: Temporal Growth of EV Adoption. Data Source: Washington State Department of Licensing (DOL), Feb. 2024.

As discussed later in Section 2, similar literature emphasized the confounding endogeneity issue associated with this approach for causal identification. Aside from omitted variables like consumer preference and awareness, technological advancements, urbanization and infrastructure, and other time-invariant characteristics, reverse causality and simultaneity are prominent concerns. The presence of more charging stations can lead to increased EV adoption as it mitigates range anxiety and enhances the convenience of EV ownership. Conversely, the increasing prevalence of EVs can drive the expansion of public charging infrastructure as businesses and state governments respond to the growing demand, creating a feedback loop (Parekh, 2023). Procuring a randomized controlled trial (RCT) on a subset of rural counties would necessitate the necessary environment to minimize selection bias while isolating the causal effect of an additional unit of public charging stations towards EV adoption. However, these experiments require substantial resources, both time and finance. Since the NEVI Form Program has progressed a third of its term, empirically analyzing these impacts would be more practical.

## 2  Literature Review

Empirical analysis on the impact of EV charging stations on EV adoption and its different subsequent variations for public accessibility and electric vehicle supply equipment (EVSE) charging types has received increasing research attention over the past few years. Understanding the magnitude of the effect of charging stations on EV adoption is particularly important given the notion of range anxiety that policymakers and business entities seek to minimize by investing in additional infrastructure units. Similar empirical studies by Sommer and Vance in 2020 employed instrumental variables to mitigate the endogeneity bias that may emerge. Their result shows that gas stations, groceries, and electrical transformers provide significant additional point estimates compared to the two-way fixed effect model (Sommer and Vance 1). However, past studies suggest that the reciprocal relationship between charging stations and EV adoptions would provide an upward bias; the instrument variable should be expected to produce a lower point estimate. The instrument variables mentioned above displayed higher marginal effects of charging stations towards EV adoptions in the same direction of the presumed bias—whereby the use of the gas station instrument variable found that subsidy for typical EV charging station leads to an increase of 0.298 BEVs per year per €1,000 subsidies on average (Sommer and Vance 2). Nevertheless, the informed revelations may be case-specific since the prevalence of fast EV charging in interstate gas stations in their region of study, Germany, is argued to be exogenous, for which the US does not appear to present the same landscape.

Another recent study conducted by Parekh in 2023 utilized a similar instrumental variable approach on 13 US states. The paper employed gas stations that had procured incentives to construct EV charging units as its primary instrument variable. Although Parekh's results highlight positive and significant point estimates on the relationship of charging stations and EV adoption comparable to Sommer and Vance, the paper suffers from an underpowered study that limits its validity in addressing the simultaneity bias through the aforementioned instrument variable approach. Consequently, this paper attempts to address the simultaneity bias issue of previous studies by employing the NEVI Formula Program as a potential instrumental variable on the relationship of charging stations towards EV adoption using the latest data available.

## 3 Data & Methodology

### 3.1 EV registration and charging infrastructure

The data analyzed in this study utilized a cleaned version of the newly updated number of monthly electric vehicle registrations from April 2017 to September 2023. The registration data, which was compiled by the Washington State Department of Licensing (DOL), integrates National Highway Traffic Safety Administration (NHTSA) data and the Environmental Protection Agency (EPA) fuel efficiency ratings with DOL titling to produce a comprehensive panel of EV registrations of 310 counties across 50 states. The dataset provides registration information for BEVs, PHEVs, as well as non-electric vehicles (non-EVs). Due to the lack of panel variation in a select few states (listed in section 3.3), the study will exclude five states with fewer than two counties in the EV registration data. Moreover, a balanced panel could not be achieved without imputing missing values. Consequently, accounting for other missing values, approximately 1,500 observations were dropped—procuring a clean panel data of 18,643 EV registration observations (approximately 98.3% of the raw dataset).

### 3.2 NEVI Formula Program

The NEVI formula program data is distributed directly by the Department of Transportation's (DOT) Federal Highway Administration (FHWA). The funding estimates for the first two years of the 5-year NEVI plan have recently been updated to reflect the actual distribution each state receives—data on the last three years remain an estimate until further notice. This dataset includes an overhead estimate on the allocation and usage of funds for the development and maintenance of EV charging stations across various states.

### 3.3 Supplementary control variables and summary statistic overview

State and county-level population estimates from the US Census are control variables. The selected data contains the aggregate estimate of populations within each county per year. Based on the findings by Sommer and Vance, average retail gasoline prices are also integrated as additional control variables to improve the point estimate. Data on the fluctuation of retail gasoline prices are taken from the US Energy Information Administration. The summary statistic for the estimation sample is provided below.

Table I. Summary statistics of the estimation sample

| Statistic | Mean | St. Dev. | Min | Max | N |
|---|---|---|---|---|---|
| Uptake of battery electric vehicles (#) | 235.165 | 2,341.016 | 0 | 70,550 | 18,643 |
| Uptake of plug-in hybrids (#) | 87.135 | 668.610 | 0 | 17,087 | 18,643 |
| Uptake of non-electric vehicles (#) | 27,606.720 | 111,888.300 | 0 | 1,399,869 | 18,643 |
| Public charging stations (#) | 70.051 | 207.351 | 0 | 3,987 | 18,643 |
| NEVI fundings ( amount) | 51,937,590.000 | 412,408,225.000 | 0 | 5,616,208,254 | 18,643 |
| County level population (#) | 578,809.600 | 984,680.900 | 2,215 | 10,103,711 | 18,643 |
| State level population (#) | 12,281,991.000 | 10,543,945.000 | 581,381 | 39,512,223 | 18,643 |
| NEVI funding round | 0.494 | 0.765 | 0 | 2 | 18,643 |
| Avg. retail gasoline price ( per gallon) | 3.059 | 0.660 | 1.938 | 5.032 | 18,643 |

Note: NEVI fundings are distributed to each state at the beginning of year 2022 and 2023.

### 3.4 Estimation strategy

Two-way fixed effect regression models (FE) will be used as specified in the estimation equation below.

$$ev_{it} = \beta_o + \beta_1 public\_evi_{it} + \beta_2 x_{it} + \theta_i + \gamma_t + \epsilon_{it}$$

Where $ev_{it}$ denotes the number of BEVs or PHEVs within a given county $i$ in period $t$, $public\_ev_{it}$ encapsulates the number of public EV charging stations, $x_{it}$ represents time-varying control variables, and $\beta$ measures the average estimate coefficients. The estimation equation will also include state-level fixed effects to control for time-invariant unobserved characteristics denoted by $\theta_i$ as well as year-month fixed effects denoted by $\gamma_t$ to control for time-varying unobserved characteristics.

The inclusion of state-level fixed effects is used to combat heterogeneity that may come from state specific factors. These sources of bias would include the advancement of EV technology, supplementary policy and incentives, the level of transportation infrastructure, as well as the regional demographic income-level—therefore allowing regression models to single out the effect of public charging stations of the same state across time. Moreover, the inclusion of the second fixed-effect, year-month, would exclude the influence of factors that changes over-time that do so uniformly across all states. These factors would include shift in public perceptions toward EVs, seasonal patterns that affects the usage and accessibility of charging stations, as well as changes in the national economy. Consequently, the employment of two-way fixed effects will enable the model to isolate the causal relationship of an increase in public charging stations towards EV adoption.

Nevertheless, one of the required identifying assumptions for the estimation method above is the absence of simultaneity—where the presence of public charging stations has a reciprocal relationship with the rate of EV adoption. This study attempts to circumvent this potential source of endogeneity using an instrumental variable approach through two-stage least squared (2SLS) of NEVI fundings towards public charging station.

### 3.5 Instrumental Variable Justification

For NEVI fundings to be an appropriate instrumental variable to mitigate simultaneous determination, or at least, significantly reduce the magnitude of the bias, it must adhere to the principles of relevance and exogenous. By construction, the objectives of the NEVI program are to allocates state fundings in procuring wider reach of charging stations regardless of the direct regional impact it has towards EV adoption. This is because the 50-mile proximity requirement of building new public charging stations limits the construction on densely targeted markets (California Energy Commission, n.d.)—emphasizing on scope-of-reach rather than density. And given that the funding is primarily used to construct public charging stations, the funding's relationship towards the public facility would be significant.

Moreover, the construction of new public charging station would not be reflected immediately after fundings are received. Although, the dispersion of funding is made when states have delivered their construction plans, the installation, testing and commissioning, as well launching would require 6 months at best for new public charging stations to be fully operational (SparkCharge, n.d.). Therefore, the estimation models will also utilize a lag outcome to see if there are any delayed effect of EV adoption.

# 4 Results

Analyzing the impact of BEV and PHEV uptake that ignores heterogeneity, Table A1. and Table 2. highlight how public charging stations pose a positive effect on both types of EV—an increase in the public facility produces an average increase of approximately 0.74 and 0.132 units of BEVs and PHEVs, respectively. The revelation that PHEVs would procure a lower estimate aligns well with other empirical studies, given that PHEVs rely less often on charging facilities due to their hybrid construction. When compared to other models that incorporate a delayed response towards EV adoption, the point estimate remains the same—this finding would remain valid for all models, including the two-stage least square model. Though the estimated direction and magnitude are comparable to previous studies, the model does not produce any significant result in the point estimates of public charging stations towards EV adoption. When looking at the different types of EV vehicle categories, trucks or similar large vehicles produce a massive increase in adoption in the presence of an additional unit of charging station, which is more prominent and statistically significant for BEVs. It could be theorized that increasing the range associated with more public facilities encourages truckers to transition towards EV complements.

| Dependent Variables: | BEVs | PHEVs | BEVs | PHEVs | BEVs | PHEVs |
|---|---|---|---|---|---|---|
| | FE (w/o controls) | | FE (w/ controls) | | FE (w/ 6-month lag) | |
| Model: | (1) | (2) | (3) | (4) | (5) | (6) |
| *Variables* | | | | | | |
| Public charging station | 2.555 | 0.7024 | 0.7396 | 0.1326 | 0.7396 | 0.1326 |
| | (3.137) | (0.8584) | (0.9882) | (0.1802) | (0.9882) | (0.1802) |
| Non-EVs | | | 0.0174*** | 0.0055*** | 0.0174*** | 0.0055*** |
| | | | (0.0003) | ($6.12 \times 10^{-5}$) | (0.0003) | ($6.12 \times 10^{-5}$) |
| Type: Truck | | | 256.6*** | 2.169 | 256.6*** | 2.169 |
| | | | (18.80) | (4.812) | (18.80) | (4.812) |
| State populations | | | $6.52 \times 10^{-5}$ | $1.03 \times 10^{-5}$ | $6.52 \times 10^{-5}$ | $1.03 \times 10^{-5}$ |
| | | | (0.0002) | ($4.97 \times 10^{-5}$) | (0.0002) | ($4.97 \times 10^{-5}$) |
| *Fixed-effects* | | | | | | |
| State FE | Yes | Yes | Yes | Yes | Yes | Yes |
| Year-month FE | Yes | Yes | Yes | Yes | Yes | Yes |
| *Fit statistics* | | | | | | |
| Observations | 18,643 | 18,643 | 18,643 | 18,643 | 18,643 | 18,643 |
| $R^2$ | 0.06402 | 0.07312 | 0.64127 | 0.78392 | 0.64127 | 0.78392 |
| Within $R^2$ | 0.04131 | 0.03874 | 0.63256 | 0.77591 | 0.63256 | 0.77591 |

*Clustered (State FE) standard-errors in parentheses*
*Signif. Codes: ***: 0.01, **: 0.05, *: 0.1*
Table 1: Fixed effect models on the uptake of EVs
Notes: Avg. retail gas price ($) was removed due to colinearity

| Dependent Variables: | Public charging station First stage | BEVs | PHEVs | BEVs | PHEVs |
|---|---|---|---|---|---|
| | | Second stage | | Second stage (w/ 6-month lag) | |
| Model: | (1) | (2) | (3) | (4) | (5) |
| *Variables* | | | | | |
| NEVI funding ($) | $2.24 \times 10^{-8}$*** | | | | |
| | $(3.9 \times 10^{-9})$ | | | | |
| Non-EVs | 0.0003*** | 0.0182*** | 0.0056*** | 0.0182*** | 0.0056*** |
| | $(4.62 \times 10^{-6})$ | (0.0003) | $(7.5 \times 10^{-5})$ | (0.0003) | $(7.5 \times 10^{-5})$ |
| Type: Truck | 37.00*** | 338.7*** | 20.43** | 338.7*** | 20.43** |
| | (5.121) | (49.13) | (9.120) | (49.13) | (9.120) |
| State populations | -0.0001 | -0.0002 | $-4.17 \times 10^{-5}$ | -0.0002 | $-4.17 \times 10^{-5}$ |
| | (0.0001) | (0.0002) | $(4.49 \times 10^{-5})$ | (0.0002) | $(4.49 \times 10^{-5})$ |
| Public charging station (fitted) | | -1.478 | -0.3609* | -1.478 | -0.3609* |
| | | (0.8840) | (0.2138) | (0.8840) | (0.2138) |
| *Fixed-effects* | | | | | |
| State FE | Yes | Yes | Yes | Yes | Yes |
| Year-month FE | Yes | Yes | Yes | Yes | Yes |
| *Fit statistics* | | | | | |
| Observations | 18,643 | 18,643 | 18,643 | 18,643 | 18,643 |
| $R^2$ | 0.25281 | 0.63807 | 0.78267 | 0.63807 | 0.78267 |
| Within $R^2$ | 0.05095 | 0.62929 | 0.77461 | 0.62929 | 0.77461 |

*Clustered (State FE) standard-errors in parentheses*
*Signif. Codes: \*\*\*: 0.01, \*\*: 0.05, \*: 0.1*
Table 3: Two-stage least squares (2SLS) models on the uptake of EVs
Notes: Avg. retail gas price ($) was removed due to colinearity

Nonetheless, a peculiar result transpires when regressing the impact of public charging stations through NEVI funding as an instrument variable, as seen in how the point estimate flips towards a converse relationship in Table 3. As hypothesized, the NEVI funding produces a highly significant effect on public charging stations—however, the fit resulted in a decrease in approximately 1.48 and 0.36 units of BEVs and PHEVs, respectively, for an increase in one unit of the public facility. To put it in perspective, an increase in an additional 3 units of public charging stations, which, per the model estimate, procure an average cost of $133,928.55, would lead to a loss of one unit of PHEVs. This development was unexpected and counterintuitive to the program's objective. Though the estimate only indicates statistical significance towards PHEVs and not BEVs, these are striking revelations that may have extensive implications for the effectiveness of the NEVI program.

When comparing the results from Sommer's and Vance's paper in Germany, their study with 12,000 units of observation provides a statistically significant enough result to accept a small effect size within a 0.007 and 0.087 standard deviation for their fast charger's instrumental variable model (Sommers and Vince). This study procured a significant result with a higher standard deviation of 0.2138. Using a power plot as seen in the appendix, a randomized controlled trial (RCT) prospective study that distributes equal treatment and control for a study with 18,643 observations and a power value of 0.8, the study would have achieved a minimum detectable effect (MDE) of approximately 0.29 standard deviations. Consequently, on the grounds of the discussed MDE and results gathered from similar studies, the study has enough statistical power to detect a null effect.

| Dependent Variables: | BEVs | PHEVs |
|---|---|---|
| | Reduced Form | |
| Model: | (1) | (2) |
| *Variables* | | |
| Constant | 236.2*** | 87.62*** |
| | (17.28) | (4.936) |
| NEVI funding ($) | $-1.96 \times 10^{-8}$ | $-9.43 \times 10^{-9}$ |
| | $(4.16 \times 10^{-8})$ | $(1.19 \times 10^{-8})$ |
| *Fit statistics* | | |
| Observations | 18,643 | 18,643 |
| $R^2$ | $1.19 \times 10^{-5}$ | $3.38 \times 10^{-5}$ |
| Adjusted $R^2$ | $-4.18 \times 10^{-5}$ | $-1.98 \times 10^{-5}$ |

*IID standard-errors in parentheses*
*Signif. Codes: \*\*\*: 0.01, \*\*: 0.05, \*: 0.1*
Table 4: Reduced-form results
Notes: Correlation of funding (IV) w/o controls

Table 4. demonstrates the reduced form of the instrumental variable regression on NEVI funding towards EV adoption. The point estimate does not indicate any statistical significance for BEVs and PHEVs. Although this regression does not signify the presence of simultaneity nor confirm the non-significance presence of its heterogeneity, the estimates' direction implies the antagonistic relationship in an increased unit of NEVI funding towards EV adoption.

## 5 Limitation

Several drawbacks in the model specification and the use of data need to be highlighted: Key control variables like average retail gas prices could not be utilized in any of the performed estimating models due to collinearity. Although fixed effects and instrumental variables are the most effective approach to deal with endogeneity—similar studies highlight how this variable provides more accurate results on their point estimates, which may flip the direction in the effect of public charging stations towards EV adoption (Narasimhan & Johnson, 2018). Moreover, given that the NEVI program is still in its early efforts and that future estimates for funding are tentative, the actual effect of the program might not be reflected immediately and is not within the reach of today's public data. More importantly, the exogenous nature of the NEVI program as a potential instrument variable of public charging stations towards EV adoption hinges on the verity that the implementation and redistribution of those funds aligned with the strict requirements they set, as highlighted previously.

## 6 Future research consideration

On the same note, another potential instrument variable is realized, which could provide a more transparent pathway for causal inference thanks to the introduction of the NEVI program. Due to the newly unlocked linkage of highways towards publicly accessible charging stations, a GIS approach to calculate the length and area of the county's access to highways could potentially provide a more robust approach compared to directly analyzing the difference in NEVI fundings as an instrument. Given that the construction of most highways in the US was completed way before the progressive transition of the EV industry, highways are highly exogenous to EV adoption and very closely connected to the construction of new public charging stations. Furthermore, a prospective study to accumulate higher regional variations for EV adoptions, particularly counties that are less urbanized, is linked with high cost for data collection—the discussed alternative would be the way forward for this particular region of studies.

## 7 Conclusion

Through analyzing five years of EV registration data on a subset of counties, this paper has analyzed the expected results from a fixed effect model and the unexpected findings of using NEVI funding as an instrument variable for public charging stations on EV adoption. This paper's chain of results suggests that an increase in public charging infrastructure may pose little significance in the broader context of this study. This paper can convey the policy implication from the NEVI estimates, suggesting that there may be more effective policies than incentivizing more EV adoption through increasing investment toward public EV infrastructure. This is not to say that the current trends of EV adoption are slowing down—the efficacy of public charging stations to garner more EV adoption might be too optimistic. Therefore, reducing the funding rate for the rest of the program's life might be an economically sound strategy when looking at the issue from a purely cost-benefit point of view. Conclusively, although the estimates from this study could potentially be drowned by endogeneity coming from a biased instrument variable, the direction of these estimates, as well as the statistical power that this study procures, are enough to discourage higher level of NEVI-related investments until new data would contend otherwise.

U.S. Department of Transportation (DOT). (n.d.). Electric Vehicles & Rural Transportation. Retrieved from https://www.transportation.gov/rural/ev#:~:text=The%20Federal%20Government%20has%20set,local%20and%20long%2Ddistance%20trips.

U.S. Department of Transportation (DOT). (n.d.). Individual Benefits of Rural Vehicle Electrification. Retrieved from https://www.transportation.gov/rural/ev/toolkit/ev-benefits-and-challenges/individual-benefits.

U.S. Department of Transportation (DOT) Federal Highway Administration (FHWA). (2024, February 02). Fiscal Year 2024 EV Infrastructure Deployment Plans. Retrieved from https://www.fhwa.dot.gov/environment/nevi/ev_deployment_plans/

U.S. Department of Transportation (DOT) Pipeline and Hazardous Materials Safety Administration (PHMSA). (2023, February 16). Bipartisan Infrastructure Law (BIL) / Infrastructure Investment and Jobs Act (IIJA). Retrieved from https://www.phmsa.dot.gov/legislative-mandates/bipartisan-infrastructure-law-bil-infrastructure-investment-and-jobs-act-iija#:~:text=The%20Infrastructure%20Investment%20and%20Jobs,%22new%22%20investments%20and%20programs.
13

# Appendix 1

| Dependent Variables: | BEVs | PHEVs | BEVs | PHEVs | BEVs | PHEVs |
|---|---|---|---|---|---|---|
| | FE (w/o controls) | | FE (w/ controls) | | FE (w/ reduced n-obv) | |
| Model: | (1) | (2) | (3) | (4) | (5) | (6) |
| *Variables* | | | | | | |
| Public charging station | 2.555 | 0.7024 | 0.7396 | 0.1326 | 1.417 | 0.2866 |
| | (3.137) | (0.8584) | (0.9882) | (0.1802) | (1.788) | (0.3631) |
| Non-EVs | | | 0.0174*** | 0.0055*** | 0.0176*** | 0.0055*** |
| | | | (0.0003) | $(6.12 \times 10^{-5})$ | (0.0001) | $(1.67 \times 10^{-5})$ |
| Type: Truck | | | 256.6*** | 2.169 | 273.8*** | 6.083*** |
| | | | (18.80) | (4.812) | (12.66) | (1.647) |
| State populations | | | $6.52 \times 10^{-5}$ | $1.03 \times 10^{-5}$ | 0.0002 | $3.59 \times 10^{-5}$ |
| | | | (0.0002) | $(4.97 \times 10^{-5})$ | (0.0003) | $(7.46 \times 10^{-5})$ |
| County populations | | | | | -0.0002 | $-5.31 \times 10^{-5}$ |
| | | | | | (0.0003) | $(6.37 \times 10^{-5})$ |
| *Fixed-effects* | | | | | | |
| State FE | Yes | Yes | Yes | Yes | Yes | Yes |
| Year-month FE | Yes | Yes | Yes | Yes | Yes | Yes |
| *Fit statistics* | | | | | | |
| Observations | 18,643 | 18,643 | 18,643 | 18,643 | 18,643 | 18,643 |
| $R^2$ | 0.06402 | 0.07312 | 0.64127 | 0.78392 | 0.64545 | 0.78657 |
| Within $R^2$ | 0.04131 | 0.03874 | 0.63256 | 0.77591 | 0.63685 | 0.77866 |

*Clustered (State FE) standard-errors in parentheses*
*Signif. Codes: \*\*\*: 0.01, \*\*: 0.05, \*: 0.1*
Table 1: Fixed effect models on the uptake of EVs
Notes: Avg. retail gas price ($) was removed due to colinearity

# Appendix 2

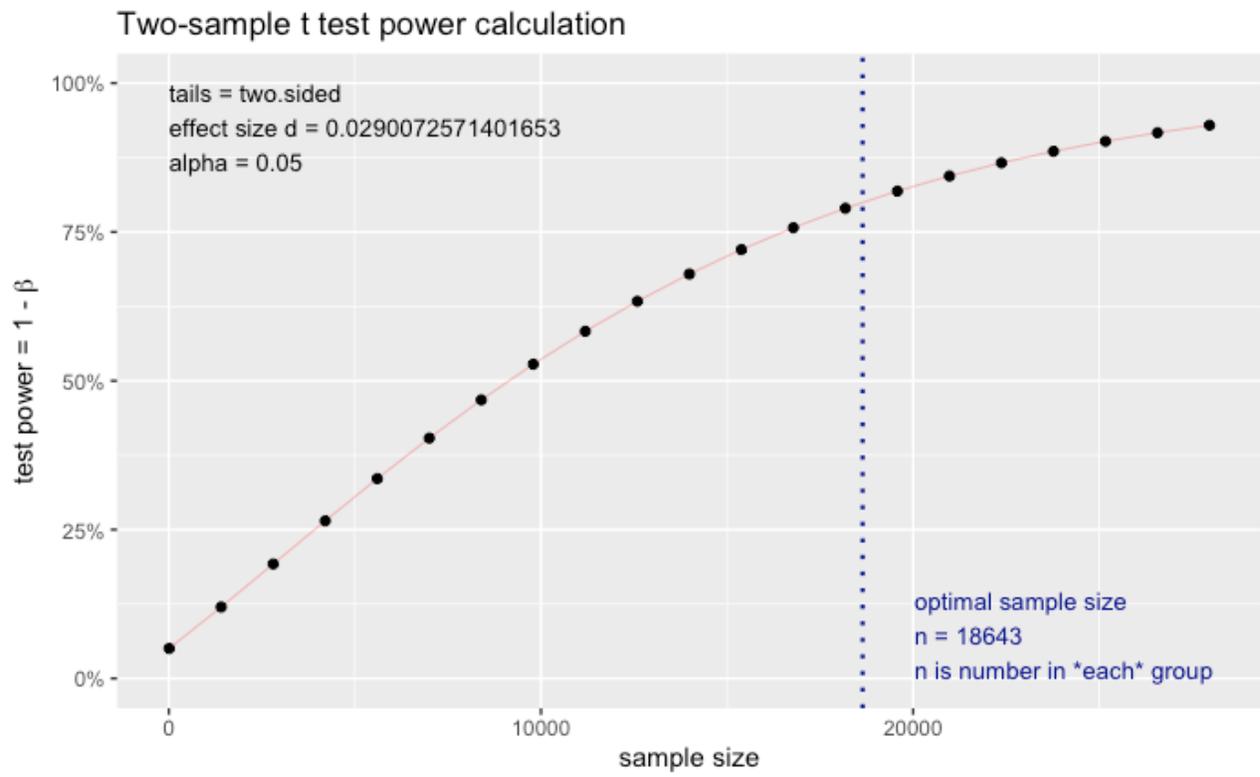